\documentclass[twocolumn,numberedappendix]{openjournal}
\usepackage{natbib}
\usepackage{graphicx}
\usepackage{graphicx,amsmath,amssymb,amstext}
\usepackage{amsbsy,amsfonts,amsthm,color}
\usepackage{caption}
\usepackage{subcaption}
\usepackage{cancel}
\usepackage{xcolor}
\usepackage{graphicx}
\definecolor{ourblue}{HTML}{118AB2}
\usepackage[colorlinks,linkcolor=ourblue,citecolor=ourblue,urlcolor=ourblue ]{hyperref}
\usepackage[title]{appendix}
 
\usepackage{siunitx}

\newcommand{\half}{\frac{1}{2}}

\newcommand{\calH}{\ensuremath{\mathcal{H}}}
\newcommand{\mathd}{\ensuremath{\mathrm{d}}}

\newcommand{\almanac}{\textsc{Almanac}}
\newcommand{\normalisingfactor}{A}
\newcommand{\ST}{{Student's~\textit{t}}}

\DeclareMathOperator{\Var}{Var}
\DeclareMathOperator{\GIG}{GIG} 
\DeclareMathOperator{\ESR}{ESR}
\DeclareMathOperator{\ESS}{ESS}
\DeclareMathOperator{\prob}{Prob}

\definecolor{ork}{rgb}{0.9,0.1,0.3}
\definecolor{grbl}{rgb}{0.3,0.6,0.7}
\definecolor{bleu}{rgb}{0,0.5,0.6}
\definecolor{armgr}{rgb}{0.29,0.33,0.13}
\definecolor{IKB}{rgb}{0,0.184,0.655}

\usepackage{soul}
\usepackage{setspace}
\usepackage{xcolor}

\usepackage{booktabs} 

\begin{document}

\journalinfo{The Open Journal of Astrophysics}
\shorttitle{Almanac: HMC sampling with bounded velocity}
\title{Almanac: Hamiltonian Monte Carlo sampling with bounded velocity}

\author{Javier Silva Lafaurie$^{1,2}$}
\author{Lorne Whiteway$^3$}
\author{Elena Sellentin$^{1,2}$}
\author{Kutay Nazli$^{1,2}$}
\author{Andrew H. Jaffe$^4$}
\author{Alan F. Heavens$^4$}
\author{Arthur Loureiro$^{4,5}$}

\affiliation{$^1$ Mathematical Institute, Leiden University, Gorlaeus Gebouw, Einsteinweg 55, NL-2333 CC Leiden, The Netherlands}
\affiliation{$^2$ Leiden Observatory, Leiden University, Gorlaeus Gebouw, Einsteinweg 55, NL-2333 CC Leiden, The Netherlands}
\affiliation{$^3$ Department of Physics and Astronomy, University College London, Gower Street, London WC1E 6BT, UK}

\affiliation{$^4$ Astrophysics Group and Imperial Centre for Inference and Cosmology (ICIC), Blackett Laboratory, Imperial College London, Prince Consort Road, London SW7 2AZ, UK}
\affiliation{$^5$ The Oskar Klein Centre, Department of Physics, Stockholm University, AlbaNova University Centre, SE-106 91 Stockholm, Sweden}

\email{j.i.silva.lafaurie@math.leidenuniv.nl}
\email{lorne.whiteway@star.ucl.ac.uk}
\email{sellentin@strw.leidenuniv.nl}
\email{nazli@strw.leidenuniv.nl}
\email{a.jaffe@imperial.ac.uk}
\email{a.heavens@imperial.ac.uk}
\email{arthur.loureiro@fysik.su.se}


\begin{abstract}
In Hamiltonian Monte Carlo sampling, the shape of the potential and the choice of the momentum distribution jointly give rise to the Hamiltonian dynamics of the sampler. An efficient sampler propagates quickly in all regions of the parameter space, so that the chain has a low autocorrelation length and the sampler has a high acceptance rate, with the goal of optimising the number of near-independent samples for given computational cost.
Standard Gaussian momentum distributions allow arbitrarily large velocities, which can lead to inefficient exploration in posteriors with ridges or funnel-like geometries. We investigate alternative momentum distributions based on relativistic and \ST{} kinetic energies, which naturally limit particle velocities and may improve robustness.
Using \almanac{}, a sampler for cosmological posterior distributions of sky maps and power spectra on the sphere, we test these alternatives in both low- and high-dimensional settings. We find that the choice of parameterization and momentum distribution can improve convergence and effective sample rate, though the achievable gains are generally modest and strongly problem-dependent, reaching up to an order of magnitude in favorable cases. Among the momentum distributions that we tested, those with moderately heavy tails achieved the best balance between efficiency and stability. These results highlight the importance of sampler design and encourage future work on adaptive and self-tuning strategies for kinetic energy parameter optimization in high-dimensional settings.
\end{abstract}

\maketitle

\section{Introduction}
\label{sec:introduction}

Markov Chain Monte Carlo (MCMC) algorithms can be used to sample probability distributions of almost arbitrary shape. Within MCMC, the Hamilton Monte Carlo (HMC) algorithm scales well for many millions of parameters \citep{2017arXiv170102434B, hoffman2011nouturnsampleradaptivelysetting}. Its fast exploration of high-dimensional parameter spaces, which relies on knowledge of the gradient of the target distribution, and the resulting rapid convergence of the sampler arise from the deterministic, measure-preserving dynamics of Hamiltonian systems \citep{2017arXiv170102434B}. HMC has a cost: the parameter space must be augmented with momentum variables. The dynamical equations (i.e. Hamilton's equations) used by HMC usually cannot be solved exactly, forcing the use of numerical integrators (with Metropolis steps to correct for discretization errors) \citep{2011hmcm.book..113N}. Most MCMC algorithms must be tuned \citep{hoffman2011nouturnsampleradaptivelysetting}; in HMC, this tuning could include the selection of a suitable probability distribution for the momentum variables (this can be viewed as defining the \textit{kinetic energy} of the system), as well as appropriately selecting the hyperparameters of the numerical integrator.

This paper focuses on some criteria to select and test the momentum distribution. If the potential energy is quadratic, so that the probability distribution is Gaussian with, say, covariance matrix $C \equiv M^{-1}$, then it is conventional wisdom that defining the kinetic energy to be $K(p) = \frac{1}{2} p^T M p$ gives efficient sampling \citep[see e.g.][Remark~2]{livingstone2019kinetic, 2011hmcm.book..113N}. However, no such recommendation exists for distributions that are not Gaussian.

For example, \almanac{} \citep{Loureiro_2023,Sellentin_2023} uses an HMC sampler to sample from a Bayesian hierarchical model to infer the posterior distribution of sky maps and power spectra, as constrained by masked and noisy spin-weight $0$ and spin-weight $2$ data on the sphere. \almanac{}'s posterior distribution exhibits a funnel under regimes with high signal-to-noise, with power law tails in particular for low multipoles. The quadratic kinetic energy that is ideal for a Gaussian distribution is therefore not necessarily ideal for this strongly non-Gaussian posterior distribution. Distributions with such funnel shapes are particularly challenging to sample from \citep{neal2003slice, betancourt2013hamiltonianmontecarlohierarchical}. While \cite{Sellentin_2023} addressed this challenge by transforming variables to flatten out the potential, another possibility would be to change the definition of kinetic energy.

The choice of kinetic energy affects both the distribution of the drawn momenta and the Hamiltonian dynamics, and therefore the impact of different choices for the kinetic energy is not easily predicted. Performance of a sampler depends (amongst other things) on the step size (i.e. the quantum of time within the discrete numerical integrator). If the probability distribution has a funnel, or sharp ridges, then smaller step sizes allow the sampler to better follow the steep gradients along the ridge \citep{2011hmcm.book..113N, betancourt2017geometric}. However, smaller step sizes drive up the computational cost. \cite{livingstone2019kinetic} have shown an example where changing from a Gaussian kinetic energy to a relativistic kinetic energy creates a sampler that is more tolerant of a wider range of step sizes. 

This paper studies two non-Gaussian kinetic energy definitions: relativistic kinetic energy and a kinetic energy definition for which the momentum has a \ST{} distribution. Both give velocity distributions that are bounded above; we study the resulting impact on the efficiency of sampling for the \almanac{} posterior.

This paper is organised as follows:
Section~\ref{sec:general_theory_of_HMC} gives the general theory of HMC, while 
Section~\ref{sec:behaviour_of_discretized_leapfrog} discusses more detailed aspects of implementing HMC using the leapfrog algorithm. 
Section~\ref{sec:kinetic_energies_with_bounded_velocity} describes the alternative kinetic energy definitions that we consider, and Section~\ref{sec:sampling_from_momentum_distributions} describes how to sample from the corresponding momentum distributions.
Section~\ref{sec:quantifying_sampler_performance} discusses how to quantify the performance of a sampler.
We finish with a description of our numerical experiments in Section~\ref{sec:almanac_experiments}, the results of these experiments in Section~\ref{sec:results}, and a discussion in Section~\ref{sec:discussion}.

\section{General theory of HMC}
\label{sec:general_theory_of_HMC}

Here is a summary of the HMC algorithm \citep{2001-Hanson-HMC, 2007-Hajian-HMC, 2011hmcm.book..113N}. Assume we wish to sample from some probability density $P$. Define the potential energy $U$ to be the negative logarithm of the density: $U(q) = -\log P(q)$, where $q$ is a point in the parameter space. Introduce a momentum variable $p$ and choose a kinetic energy function $K(p)$ satisfying the symmetry constraint $K(p) = K(-p)$. The Hamiltonian total energy is then
\begin{equation}
    H(q,p) = U(q) + K(p).
\end{equation}
We may sample from the density $\exp(-H)$ defined on phase space $(q, p)$ by alternately `following the Hamiltonian flow' and then `resampling from the density $\exp(-K)$ on momentum space'; each of these steps individually preserves detailed balance while between them they allow all parts of phase space to be reached. Marginalising over momentum (i.e. simply discarding the momentum components from the samples) then yields samples of $P$. Further information on HMC may be found in \cite{2011hmcm.book..113N} and \cite{betancourt2017geometric}.

In the above, we have ignored constant normalising factors for the probability density functions as such factors make a constant contribution to the energy and hence have no impact on the dynamics; in later sections we will include such a factor (denoted $\normalisingfactor{}$) in the specification of the momentum probability density when it is interesting to do so.

To follow the Hamiltonian flow we must solve Hamilton's equations
\begin{align}
\begin{split}
    \dot{q} & = \frac{\partial H}{\partial p} = \frac{\mathd K}{\mathd p} ,\\
    \dot{p} & = -\frac{\partial H}{\partial q} = - \frac{\mathd U}{\mathd q}.
\end{split}
\end{align}
For solving these equations numerically, it is practical to discretize time (via a time step parameter $\epsilon$) and use the leapfrog scheme
\begin{align}
\begin{split}
    p_i (t + \epsilon/2) &= p_i(t) - \frac{\epsilon}{2} \frac{\partial U [ q(t)]}{\partial q_i} \\
    q_i(t + \epsilon) &= q_i(t) + \epsilon \frac{\partial K[p(t+\epsilon/2)]}{\partial p_i} \\
    p_i(t + \epsilon) &= p_i(t + \epsilon/2) - \frac{\epsilon}{2} \frac{\partial U [ q(t + \epsilon)]}{\partial q_i} \ .
    \label{leapfrog_scheme}
\end{split}
\end{align}
Several such updates may be concatenated to give a single trajectory. Leapfrog is useful as it is both time-symmetric and symplectic (i.e. it preserves phase space volume). However it is an approximation and a leapfrog trajectory typically will not conserve total energy exactly; for detailed balance a further Metropolis accept/reject step is therefore needed at the end of each trajectory. The acceptance rate for this step may be made arbitrarily close to $100\%$ (at the expense of increased compute time) by decreasing $\epsilon$ while increasing the number of leapfrog updates in a trajectory (as this will decrease each trajectory's energy error).

\section{Behaviour of discretized leapfrog}
\label{sec:behaviour_of_discretized_leapfrog}

Consider how the leapfrog integrator updates the position $q$. Let $q(0)$ and $p(0)$ be, respectively, the position and the (randomly selected) momentum at the beginning of the leapfrog loop. After a single leapfrog update the change in sampler position will be 
\begin{equation}
    q(\epsilon) - q(0) = \epsilon \nabla K\left[  p(0) - \frac{\epsilon}{2} \nabla U[q(0)]  \right],
    \label{posupdate_new}
\end{equation}
which will be large when $\nabla K$ is. Large position changes are not necessarily bad; indeed, the whole point of HMC is to propose new positions that are distant from the starting position but that are nevertheless likely to be accepted. However, large position changes are more likely than small position updates to be associated with non-conservation of energy and hence possibly low acceptance rates.

\cite{livingstone2019kinetic} show that in many cases the position dynamics are governed by the behaviour of the \textit{composite gradient} function $\nabla K \circ \nabla U$. It is clear from Eq.~\ref{posupdate_new} that this function must play some role, particularly when $||p(0)||$ is negligible compared to $||\nabla U||$. The $-\epsilon/2$ factor is handled via $\nabla K$ being odd (since $K$ is assumed even), together with a homogeneity assumption that is typically satisfied in practical examples (in \cite{livingstone2019kinetic} see both Eq.~10 and Remark 1). The paper shows for example that HMC sampling of light-tailed distributions is difficult when the composite gradient grows super-linearly in its argument (there are additional technical assumptions).

As a simple example consider sampling a Gaussian distribution using Gaussian momenta; here both $U$ and $K$ will be quadratic, so $\nabla U$ and $\nabla K$, and hence the composite gradient, will be linear.

\cite{livingstone2019kinetic} also discuss the behaviour of HMC in a neighbourhood of the peak. However in our examples the high dimensionality of the parameter space results in negligible prior volume near the peak; the sampler will spend essentially all its time in the typical region\footnote{For a Gaussian distribution in $d \gg 1$ dimensions, the probability mass concentrate in a thin shell called the typical region, which is located at a radius of order $\sqrt{d}$ standard deviations from the peak. The radial width of this shell is of order $\sqrt{2}/2$, independent of $d$.} and virtually none in the vicinity of the peak \citep{2011hmcm.book..113N, betancourt2017geometric}. Therefore, we will not discuss sampler behaviour near the peak.

In this paper we consider kinetic energy definitions in which $\nabla K$, and hence the composite gradient and the velocity $\dot{q}$, have a finite upper bound. For such samplers we know that no matter how steep a gradient of the potential the sampler encounters, and no matter how extreme a starting momentum value is drawn, the sampler will always propagate at or below a certain maximum velocity. 

\begin{figure}%
    \centering
    \includegraphics[width=0.45\textwidth]{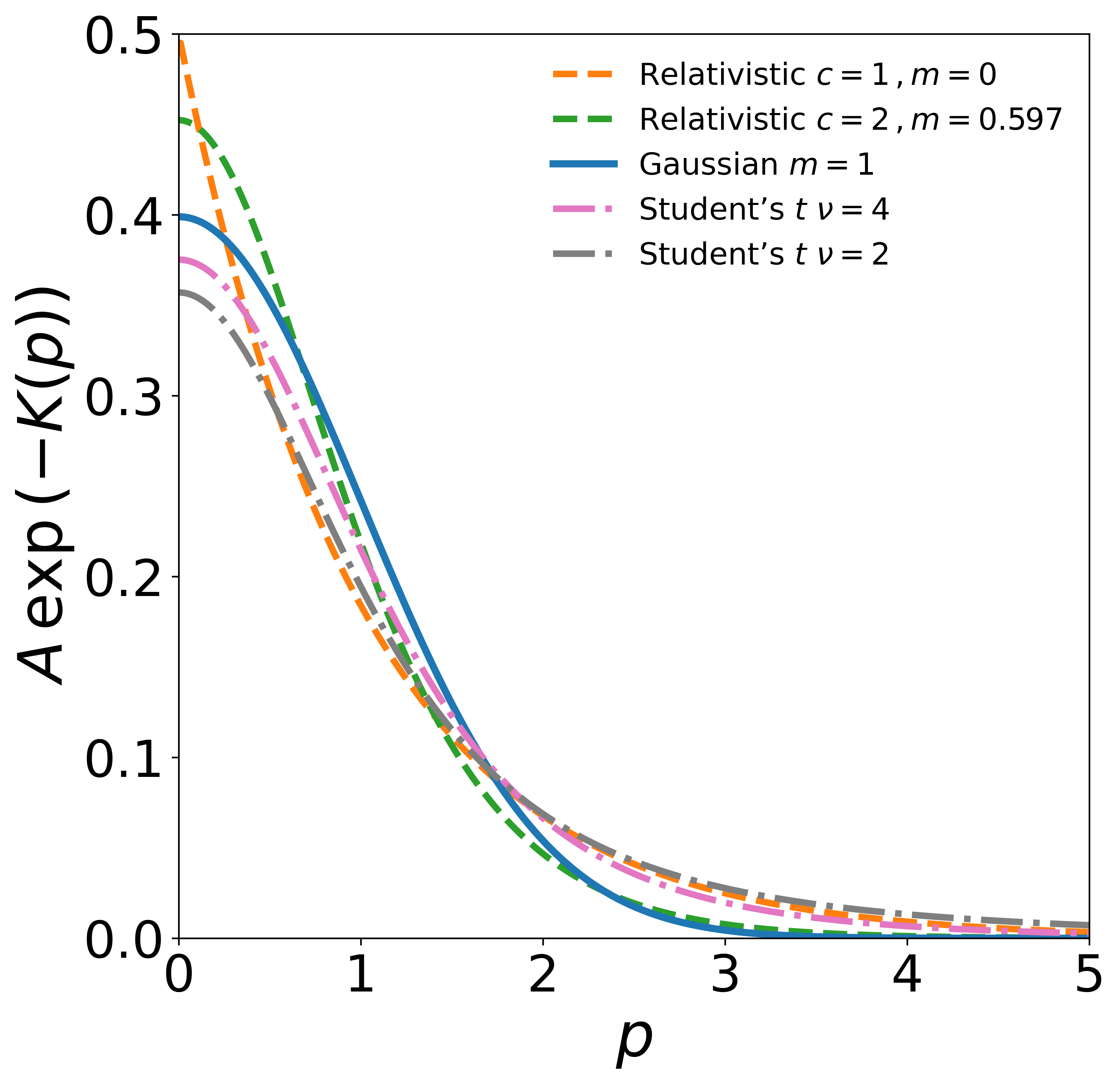}   
    \caption{Momentum proposal probability density functions in one dimension: Relativistic (orange and green dashed line), \ST{} (pink and gray dots), and Gaussian (blue solid line). The non-Gaussian distributions have fatter tails (and hence propose large momenta more often); apart from that the distributions appear similar. This visual similarity masks how strongly the non-Gaussian distributions change the Hamiltonian flow: their gradients ensure that particles cannot propagate above a maximum velocity (see Fig.~\ref{dist_v}). In each case we show the right half of a density function that is symmetric about $p=0$. The standard deviations of the plotted distributions are: $\sqrt{2}$ for the Relativistic($c=1$, $m=0$) and for the $\nu=4$ \ST{}, unity for the Relativistic($c=2$, $m=0.597$) and for the Gaussian, and undefined for the $\nu=2$ \ST{}.}
    \label{dist_p}%
\end{figure}

\begin{figure}%
    \centering
    \includegraphics[width=0.45\textwidth]{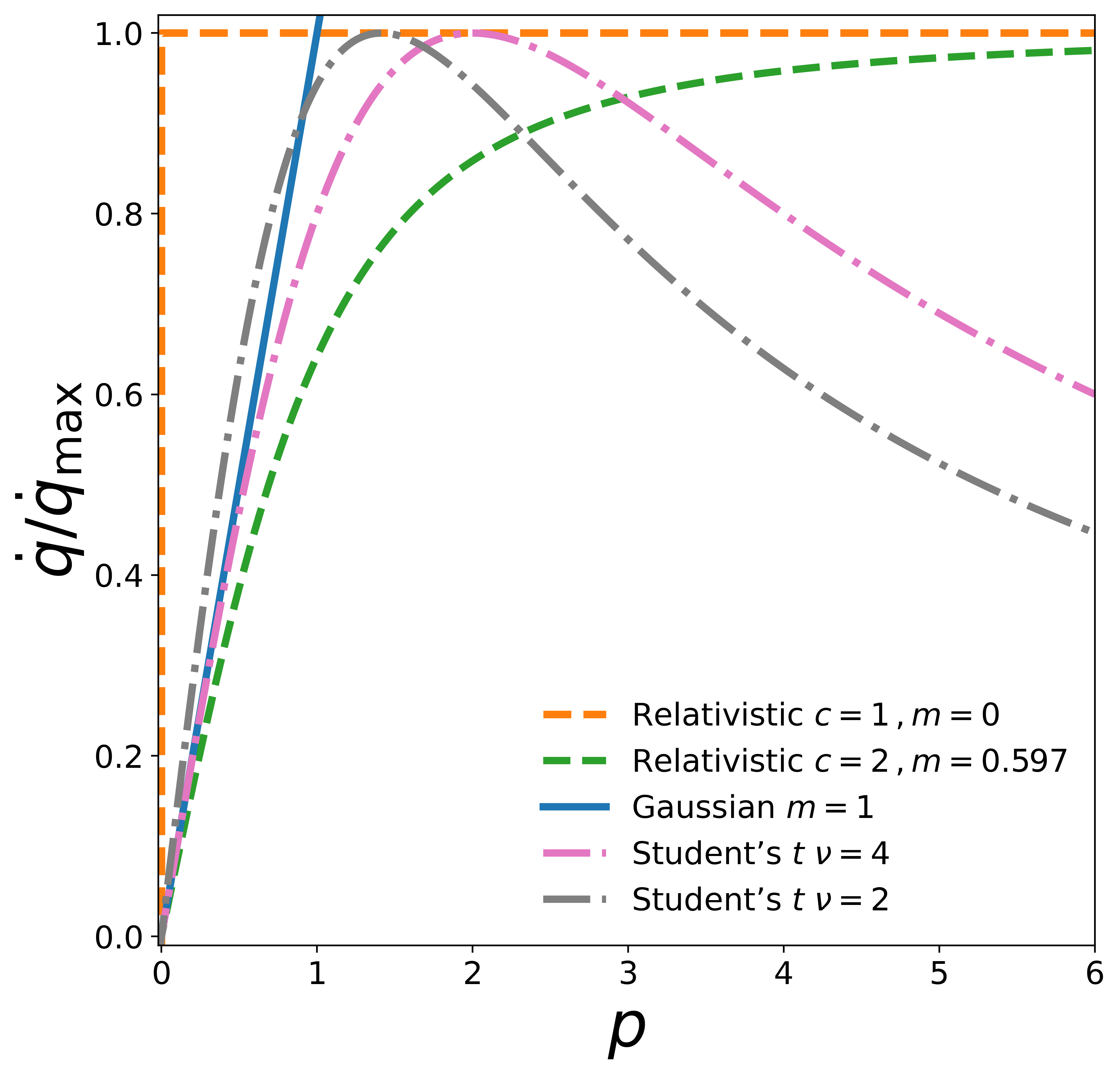}  
    \caption{Velocity $\dot{q} = p/M(p)$ as a function of $p$ for the same examples as in Fig.~\ref{dist_p}. The Gaussian distribution (with quadratic kinetic energy) has unconstrained velocity (blue solid line). For the relativistic kinetic energy, we have a maximum velocity of $c$ for a photon (orange solid line) and a massive particle (green dashed line). The mass determines the particle's behaviour near the minimum of the potential, and the maximum velocity in the outskirts. For the \ST{} kinetic energy (pink and gray dotted lines), the maximum velocity is given by Eq.~\ref{max_velocity_student}. For each non-Gaussian example, the $y\text{-axis}$ is normalised by the maximum velocity $\dot{q}_{\rm max}$. }
    \label{p_vs_v}%
\end{figure}

\begin{figure}%
    \centering
    \includegraphics[width=0.45\textwidth]{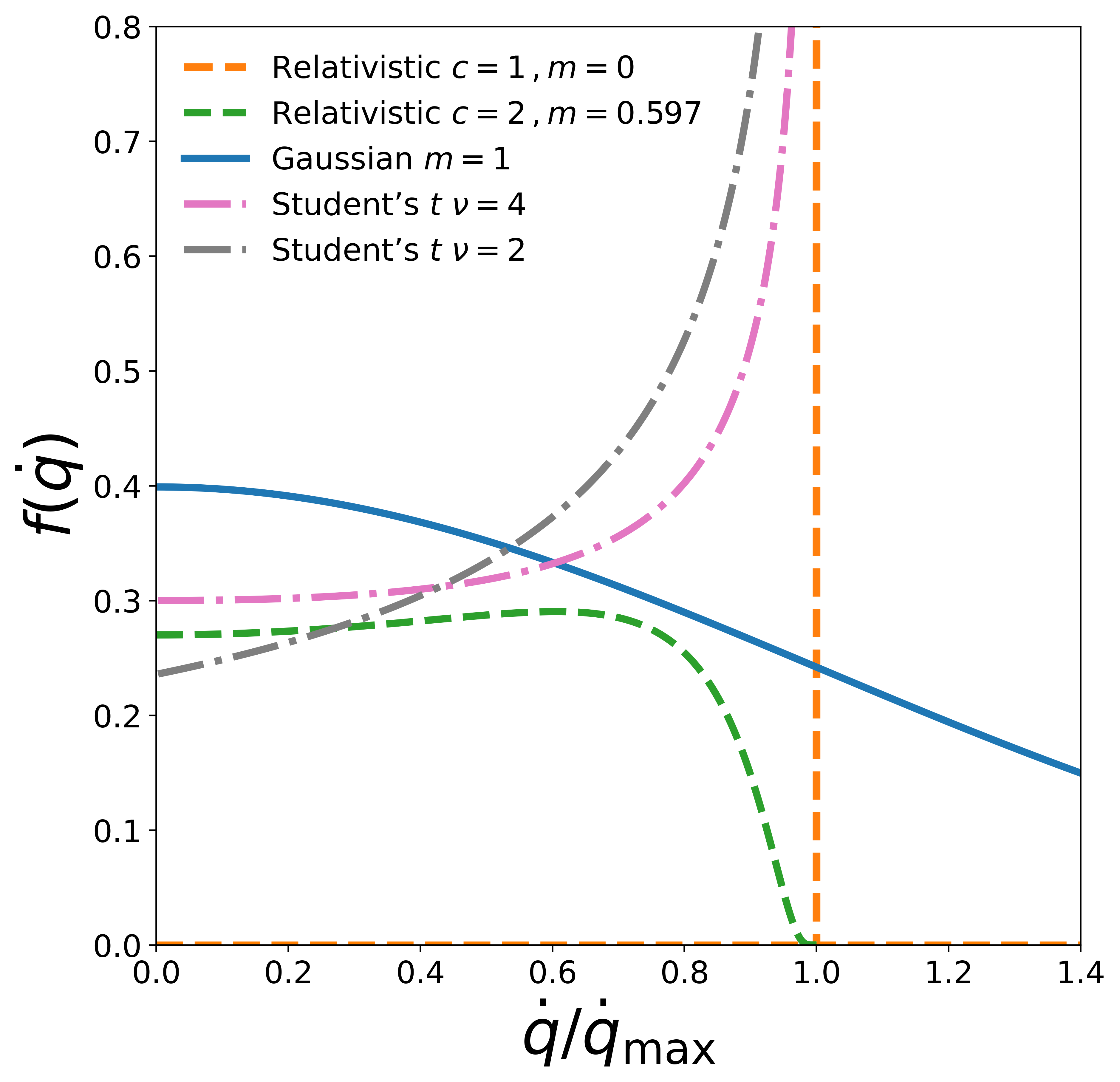}  
    \caption{Velocity proposal distributions for same examples as in Figs.~\ref{dist_p} and \ref{p_vs_v}. The Gaussian distribution (blue solid line) allows all velocities. For the relativistic kinetic energy, we observe a delta function at $\dot{q}=1$ when $m=0$ (orange dashed line) and for massive particles a distribution that proposes velocities in the range $(-1,1)$, which is less likely to propose velocities close to the maximum (green dashed line). In contrast, for the \ST{} kinetic energy (pink and gray dotted lines), it is more likely to propose velocities close to the maximum. For each non-Gaussian example, the $x\text{-axis}$ is normalised by maximum velocity $\dot{q}_{\rm max}$. In each case we show the right half of a density function that is symmetric about $\dot{q}=0$.}
    \label{dist_v}%
\end{figure}

\section{Kinetic energies with bounded velocity}
\label{sec:kinetic_energies_with_bounded_velocity}

\subsection{Relativistic}

Following \cite{lu2017relativistic}, we fix parameters $m \ge 0$ (the `rest mass') and $c > 0$ (the `speed of light'). The relativistic kinetic energy\footnote{For convenience, we define the kinetic energy to include the rest energy, which contributes only a constant offset and therefore does not affect the Hamiltonian dynamics. Note also that $p$ and $q$ denote vectors, although we suppress vector notation for clarity.} is
\begin{equation}
    K(p) = c^2\left( m^2 + \frac{|p|^2}{c^2}\right)^{\half}.
    \label{K}
\end{equation}
This is hyperbolic in $|p|$, and hence quadratic for small $|p|$ and linear for large $|p|$. Therefore, we recover Gaussian-like dynamics near the peak of the target distribution at large $|p|$ \citep{lu2017relativistic, livingstone2019kinetic}.

Such a redefinition of the kinetic energy will alter the Hamiltonian flow. Applying Eq.~(\ref{K}) gives Hamilton's velocity equation the form
\begin{equation}
    \dot{q}  = \frac{\partial H}{\partial p} = \frac{p}{ M(p) },
    \label{hamilton_velocity}
\end{equation}
where
\begin{equation}
    M(p) = \left(  m^2 + \frac{|p|^2}{c^2} \right)^\half
    \label{relativistic_mass}
\end{equation}
is the relativistic mass. We see $|\dot{q}| \le c$, with equality if and only if $m = 0$. 

\cite{lu2017relativistic} show that a separable form of the kinetic energy outperforms the multivariate distribution Eq.~(\ref{K}). The separable kinetic energy is
\begin{equation}
    K(p) = c^2\sum_i \left( m_i^2 + \frac{p_i^2}{c^2}\right)^{\half}
\end{equation}
i.e. it is a sum of one-dimensional relativistic functions over the parameter space. It is this separable form that is used in the numerical results for \almanac{} reported in this paper. We use corresponding versions of the dynamics (Eq.~\ref{hamilton_velocity}) and the relativistic mass (Eq.~\ref{relativistic_mass}) that operate component-wise. In the remainder of this section we discuss only the one-dimension relativistic distribution.

Changing the kinetic energy definition also changes the distribution $ \normalisingfactor{} \exp(-K(p))$ from which the momentum is drawn. The normalising factor is
\begin{equation}
\normalisingfactor{} = \left[ 2 m c \, K_{1}(m c^2) \right]^{-1} \, .
\end{equation}
Here and in what follows $K_{n}$ denotes the degree $n$ modified Bessel function of the second kind.

Fig.~\ref{dist_p} compares the relativistic momentum distribution and the corresponding Gaussian momentum distribution. We see that both distributions propose high and low momenta, but the relativistic distribution has a fatter tail and hence proposes a higher fraction of large momenta (considering that both distributions have similar variances). 

Fig.~\ref{p_vs_v} shows the behaviour of the velocity of different particles as a function of the momentum. The speed of light can be tuned to set the maximum velocity of the sampler, while the mass can be tuned to have good sampling performance near the peak of the distribution. Importantly, the finite speed of light will cause the sampler to propagate moderately through regions of large potential gradient $\nabla U(x)$.

The probability distribution of relativistic momentum, coupled with the relationship between momentum and velocity, induces a probability distribution of the velocity. For $0 \le \dot{q} < c$ this probability density is
\begin{equation}
    \prob(\dot{q}) = \normalisingfactor{} m c^3 (c^2-\dot{q}^2)^{-\frac{3}{2}} \exp\left[-m c^3 (c^2-\dot{q}^2)^{-\frac{1}{2}} \right]
\end{equation}
provided that $m>0$; if $m=0$ then $\prob(\dot{q})$ is a delta function with all probability mass at $\dot{q} = c$.
Fig.~\ref{dist_v} shows the probability density of $\dot{q}$ for certain examples. The density has the the following features when $m>0$:
\begin{itemize}
\item{$\prob(\dot{q})$ tends to zero as $\dot{q}$ tends to $\dot{q}_{\textrm{max}} = c$;
}
\item{The slope of the density of $\dot{q}$ is zero at $\dot{q} = 0$ (this is a consequence of $K$ being an even function, together with $\prob(\dot{q})$ being differentiable at $0$).}
\end{itemize}

\subsection{\ST{}}

Fix a parameter $\nu>0$ (called the `number of degrees of freedom', but not necessarily an integer). The \ST{} kinetic energy is
\begin{equation}
    K(p) = \frac{1+\nu}{2}\log\left( 1+\frac{|p|^2}{\nu}\right),
    \label{ST}
\end{equation}
which is quadratic for small $|p|$ and logarithmic (hence sublinear) for large $|p|$. Velocity equation Eq.~\ref{hamilton_velocity} continues to hold, but now where
\begin{equation}
    M(p) = \frac{  \nu + |p|^2 }{1 + \nu}
    \label{ST_mass}
\end{equation}
is the \ST{} mass. 

As before with the relativistic case, there is a separable version of the \ST{} kinetic energy:
\begin{equation}
    K(p) = \frac{1 + \nu}{2}\sum_i \log\left(1 + \frac{p_i^2}{\nu}\right)
\end{equation}
with corresponding componentwise velocity and \ST{} mass. As before, it is this separable version that is used in numerical examples. And as before, in the remainder of this section we discuss only the one-dimension \ST{} distribution.

Fig.~\ref{dist_p} shows the momentum probability density $\normalisingfactor{} \exp(-K(p))$ for the \ST{} kinetic energy, where here the normalising constant is
\begin{equation}
\normalisingfactor{} = \frac{\Gamma[(1+ \nu) / 2]}{\sqrt{\pi \nu} \, \Gamma(\nu / 2)} \, .
\end{equation}
As before, Fig.~\ref{p_vs_v} shows the behaviour of the velocity as a function of momentum. The maximum velocity is
\begin{equation}
    \dot{q}_{\textrm{max}} = \frac{1 + \nu}{2 \sqrt{\nu}}.
    \label{max_velocity_student}
\end{equation}
This occurs at $p^2 = \nu$; for larger $p^2$ the velocity turns over and approaches zero.

In one dimension for $0 < \dot{q} < \dot{q}_{\textrm{max}}$ the induced probability distribution of the velocity is 
\begin{equation}
    \prob(\dot{q}) = \frac{\normalisingfactor{} \nu}{\sqrt{\Delta}} \left[ \left(\frac{\dot{q} p_{-}}{1+\nu} \right)^{\alpha} + \left(\frac{\dot{q} p_{+}}{1+\nu} \right)^{\alpha} \right] .
\end{equation}
Here $\alpha = (\nu-1)/2$, $\Delta = (1+\nu)^2 - 4 \nu \dot{q}^2$, and $p_{\pm}$ are the two momenta corresponding to $\dot{q}$ via Eqs.~\ref{hamilton_velocity} and \ref{ST_mass} i.e.
\begin{equation}
    p_{\pm} = \frac{(1+\nu) \pm \sqrt{\Delta}}{2\dot{q}} \ .
\end{equation}
 Fig.~\ref{dist_v} shows the probability density of $\dot{q}$ for certain examples. The density has the the following features:
\begin{itemize}
\item{
$\prob(\dot{q})$ tends to infinity as $\dot{q}$ tends to $\dot{q}_{\textrm{max}}$;
}
\item{
If $\nu > 2$ then the slope of the density of $\dot{q}$ is zero at $\dot{q} = 0$, while if $\nu < 2$ then $\prob(\dot{q})$ has an asymptote at $\dot{q} = 0$. In the intermediate case $\nu = 2$ we have instead
    \begin{equation}
        \lim_{\dot{q} \to 0^{+}} \frac{\mathd \prob(\dot{q})}{\mathd \dot{q}} = \normalisingfactor{} \frac{2 \sqrt{2}}{9} = \frac{1}{9}\, ;
    \end{equation}
this behaviour is evident in Fig.~\ref{dist_v}. 

}
\end{itemize}

\subsection{Comparison of shapes of velocity distributions}

Fig.~\ref{dist_v} illustrates the contrasting nature of the velocity distributions arising from the relativistic and the \ST{} kinetic energies. Consider in turn:
\begin{itemize}
\item{
    Relativistic kinetic energy with $c=2$ and $m = 0.597$. (The parameters of this distribution were chosen to yield a momentum distribution with standard deviation of unity -- see Eq.~\ref{relativistic_momentum_variance}). The velocity distribution is the green dashed line in Fig.~\ref{dist_v}. The density resembles a top-hat with a cut-off at about $80\% $ of the maximum velocity; the interval $2/3 < \dot{q} / \dot{q}_{\textrm{max}} < 1$ contains only $11\% $ of the probability mass.
}
\item{
    \ST{} kinetic energy with $\nu = 4$. (The parameter yields a momentum distribution with standard deviation $\sqrt{2}$ -- see Eq.~\ref{students_t_momentum_variance}). The velocity distribution is the pink dotted line in Fig.~\ref{dist_v}. The density has more of its probability mass in the region close to the maximum velocity; the interval $2/3 < \dot{q} / \dot{q}_{\textrm{max}} < 1$ contains $48\% $ of the probability mass.
}
\end{itemize}

Thus the two kinetic energies both have bounded velocity, but differ essentially in how the velocity probability density is distributed between $0$ and the maximum velocity.

The actual value of $\dot{q}_{\textrm{max}}$, which depends on the parameters of the non-Gaussian kinetic energy, will prove to be somewhat irrelevant in our numerical examples using $\almanac{}$. This is because the velocity is degenerate with the the step-size (i.e. the $\epsilon$ parameter in Eq.~\ref{leapfrog_scheme}), and the latter is a `tuned' parameter within the \almanac{} sampler. (Specifically, for each component the step size is set to be the standard deviation of the sampled values for that component from a post-burn-in set of tuning samples, then multiplied by an overall scale factor chosen so that the sampler achieves a target acceptance rate. See \cite{Sellentin_2023} Section 5.2 for details.) The importance of the parameters in the kinetic energy definition lies therefore in setting the shape of the velocity distribution, and not in setting the absolute level of the maximum velocity. Note that as $c$ (resp. $\nu$) for the relativistic (resp. \ST{}) kinetic energy tends to infinity, we simply recover the Gaussian case. We conclude that detailed investigation of the parameter space for the kinetic energy definition is unwarranted.

\section{Sampling from momentum distributions}
\label{sec:sampling_from_momentum_distributions}

\subsection{Relativistic}

The hyperbolic distribution \citep{hyperb} is a continuous distribution with support over the entire real line. It arises as a mixture of normal distributions, and this leads to a sampling algorithm for this distribution, described below. 

The density of the hyperbolic distribution with location parameter $\mu$ and shape parameters $\alpha$, $\beta$, and $\delta$ is
\begin{multline}
   \calH(x|\mu,\alpha,\beta,\delta)  = \\
   A(\alpha,\delta,\gamma) \exp\left[ -\alpha \sqrt{\delta^2 + (x-\mu)^2} +\beta (x-\mu)\right] \, ,
\end{multline}
where $\gamma = \sqrt{\alpha^2-\beta^2}$ and
\begin{equation}
    A(\alpha,\delta,\gamma) = \frac{\gamma}{2\alpha\delta K_1(\delta\gamma)} \, .
\end{equation}

This distribution has mode
\begin{equation}
   \hat{x} = \mu + \frac{\delta\beta}{\gamma},
\end{equation}
mean
\begin{equation}
    \langle x\rangle = \mu + \frac{\delta \beta K_2(\delta\gamma)}{\gamma K_1(\delta\gamma)} \, ,
\end{equation}
and variance
\begin{equation}
    \Var(x) = \frac{\delta K_2(\delta\gamma)}{\gamma K_1(\delta\gamma)} + \frac{\beta^2\delta^2}{\gamma^2} \left[ \frac{K_3(\delta\gamma)}{K_1(\delta\gamma)} - \frac{K_2^2(\delta\gamma)}{K_1^2(\delta\gamma)} \right]
    \, .
\end{equation}

The relativistic momentum distribution is a special case of the hyperbolic distribution with
\begin{equation}
    \beta = \mu = 0, \qquad \alpha = c, \quad \textrm{and} \quad \delta = m_ic.
\end{equation}
The condition $\beta = 0$ is necessary to get $\hat{x} = \langle x\rangle = 0$, which is forced by the symmetry of the momentum distribution. As a consequence $\gamma = \alpha \ (= c)$. 
The variance of relativistic momenta then is
\begin{equation}
    \Var(p_i) = m_i \frac{K_2(m_ic^2)}{K_1(m_ic^2)},
    \label{relativistic_momentum_variance}
\end{equation}
which we use for matching the variances of Gaussian and relativistic momenta. From the variance Eq.~(\ref{relativistic_momentum_variance}) we see that a photon of zero rest mass has a momentum variance of zero; in other words it always propagates at the speed of light.
Since $(K_2/K_1)(x)$ approaches 1 when x is large but is $2/x$ when x is near zero, the variance equals the rest mass when $E \equiv m_i c^2$ is large, but is bounded below by $2/c^2$ when $E$ is near zero.

A draw from the relativistic distribution may be obtained by drawing from a mean zero Gaussian distribution whose variance is itself a draw from the Generalised Inverse Gaussian (GIG) distribution. See \cite{Dagpunar01011989} for an efficient algorithm for drawing samples from the GIG distribution using rejection sampling; in the notation of that article, we require $\GIG(\lambda=1, \psi = c^2, \chi = m^2 c^2)$.

\subsection{\ST{}}

The \ST{} distribution with $\nu > 2$ has variance
\begin{equation}
    \Var(p_i) = \frac{\nu}{\nu - 2}.
    \label{students_t_momentum_variance}
\end{equation}

Sampling from the \ST{} distribution may be done using the method of \cite{bailey1994polar}. See \cite{ShawStudentsT} for an efficient implementation.

\section{Quantifying Sampler Performance}
\label{sec:quantifying_sampler_performance}

The performance of an MCMC sampler can be quantified as the number of \emph{near-independent} samples drawn from the target distribution for a given computational cost (MCMC chains are autocorrelated, and so the number of near-independent samples will be smaller than the actual number of samples $N$). Although never strictly independent, we quantify the number of effectively independent samples using the \emph{Effective Sample Size} ($\ESS$), defined to be
\begin{equation}
    \ESS = \frac{N}{\hat{\tau}}
\end{equation}
where $\hat{\tau}$ is a correlation length estimator \citep[Eq.~17]{corr_length_estimator}. By construction, $1 \leq \ESS \leq N$.
We may similarly define $\hat{\tau}_i$ and $\ESS_i$ to be the correlation length and the Effective Sample Size of the $i$th dimension. 

For a complicated non-Gaussian distribution, $\hat{\tau}_i$ can be highly variable as a function of $i$. This typically creates an MCMC chain in which some problematic dimensions of the sampler have a much higher correlation length than the majority of variables, propagating also this behaviour to the Effective Sample Size $\ESS_i$ \citep{betancourt2013hamiltonianmontecarlohierarchical, 2011hmcm.book..113N}. In later sections, we analyze how parameterization and kinetic-energy definitions influence these behaviors.

Finally we need to account for the computational cost of the sampling algorithm, so for a particular computer configuration define the Effective Sample Rate for each parameter $i$ as 
\begin{equation}
    \ESR_{i}=\frac{\ESS_{i}}{{t_{\rm total}}}\,,
\label{def_ESR}
\end{equation}
where $t_{\rm total}$ is the time for the sampler to calculate the $N$ samples on a given architecture. This provides a direct, hardware-dependent measure of sampling efficiency. The product $(\ESR\times{\hat\tau)}$ is equal to the time to produce a single sample, and will usually be roughly constant for runs on a single hardware environment. We will use $\ESR{}$ in the following sections to compare performance across parameterizations and kinetic-energy definitions.

\section{Almanac Experiments}
\label{sec:almanac_experiments}

We use the HMC sampler of \almanac{} \citep{Loureiro_2023,Sellentin_2023} to test and check the performance of the alternative kinetic energy definitions described in Section~\ref{sec:kinetic_energies_with_bounded_velocity}. \almanac{} samples the posterior distribution of spherical harmonic coefficients (sky maps) and power spectra, as constrained by input data on the sphere (masked, noisy, and of spin-weight 0 and/or 2). This type of problem appears in cosmological experiments that study the cosmic microwave background, weak lensing, and galaxy clustering \citep[see e.g.][]{planck2018parameters,
wmap2013nineyear, kids1000_2021, des_y1_2017, desi_fullshape_2024}.

\almanac{} uses a Bayesian hierarchical model: the posterior contains terms describing the relationship between data and sky map, between sky map and power spectrum, and between power spectrum and its prior. Our ultimate target is the marginal distribution of the power spectrum; the sky map parameters are nuisance parameters \citep{Jewell_2004, Wandelt_2004}. However, the latter are extremely high-dimensional, and so HMC and Gibbs sampling are the only practical options for sampling that satisfies detailed balance \citep{Eriksen, 2011hmcm.book..113N}. \almanac{} can operate using two different parameterizations. The parameterizations differ both in how they represent the sky maps (although in both cases this is based on a representation of the sky map using spherical harmonic functions of the appropriate spin-weight) and in how they represent the elements of the power spectra (recall that \almanac{} can model several correlated fields, so in general for each multipole $\ell$ the power spectrum is a covariance matrix $C_{\ell}$, not a single number) \citep{Sellentin_2023, Loureiro_2023}.

\begin{itemize}
  \item \emph{Classic}: We take as parameters the entries of the matrix logarithm of $C_{\ell}$ (for each multipole $\ell$), as well as the (real and imaginary parts of the) spherical harmonic coefficients of the maps.
  \item \emph{Cholesky}: Begin by taking the Cholesky decomposition $L_{\ell}$ of each $C_{\ell}$ (so that $L_{\ell}$ is lower-triangular and $L_{\ell} L_{\ell}^{T} = C_{\ell}$; $L_{\ell}$ is an `effective square root' of $C_{\ell}$). We take as parameters the entries of each $L_{\ell}$ (or, for the diagonal entries, the logarithm of this entry). For the sky maps we take as parameters the spherical harmonic coefficients of maps, but (for each $\ell$) normalised by $L_{\ell}$ (so as to have parameters with unit variance).
\end{itemize}
Both parameterizations have the useful property that no part of the parameter space is forbidden. With the Classic parameterization a funnel may appear in high signal-to-noise regimes; the Cholesky parameterization was introduced to partially address this problem. Additionally, with the Classic parameterization the calculation of the derivative of the potential energy requires us to calculate the second derivative of `matrix exp' with respect to the entries of the input matrix (Eq.~C11 in \cite{Sellentin_2023}); this calculation is time-consuming and makes the Classic parameterization intrinsically slower than the Cholesky parameterization. We test different combinations of parametrizations and kinetic energy definitions to corroborate these observations.

We perform three sets of experiments: one with a mid-dimensional parameter space (the mid-dim set), one with a high-dimensional parameter space (the high-dim set), and one with a substantially lower-dimensional parameter space (the lower-dim set). In each set we do several \almanac{} runs, varying the parameterization and the kinetic energy definition used. Within each set, the system conditions (computer processor, number of nodes and cores, etc.) are held constant: we use a High-Memory Node (AMD EPYC 7662, 2.0GHz, 128 cores), and we parallelize with 20 cores (for the mid-dimensional and lower-dimensional sets) or 30 cores (for the high-dimensional set). For each \almanac{} run we note the time $t_{\rm total}$ required to produce $N=100\,000$ samples. This $N$ was chosen to reach convergence in the power spectra variables; we check the convergence using Hanson's statistic and the Fraction of Missing Information (FMI) as in \cite{Sellentin_2023}. FMI is computed as in \cite{2016arXiv160400695B, 2017arXiv170102434B}; it quantifies the efficiency of the sampler by measuring how well the marginal energy distribution matches the conditional energy distribution, assessing whether the sampler is properly exploring the full range of total energy submanifolds within the phase space. Finally, we use Eq~\ref{def_ESR} to compute the Effective Sample Rate for each parameter (noting that it is the power spectrum parameters in which we are most interested, as they are the ones of cosmological interest).

Each \almanac{} run progresses as follows (see \cite{Sellentin_2023} for details): a burn-in period of $20\,000$ samples to find the typical region, then $20\,000$ samples used to set the time step $\epsilon$ for each parameter; then a further $5\,000$ samples for `leapfrog tuning' (used to scale all the step sizes to achieve an acceptance rate of $0.75$), and finally $100\,000$ samples to be used in the analysis.

\begin{table*}[t]
    \centering
    \begin{tabular}{c c c c c c c}
     \toprule
        Set & Momentum distribution & Parameters & Parameterization & FMI & Averaged $\hat{\tau}$ & Averaged $\ESR{} ($\qty{}{\per\second}$)$ \\
        \midrule
        Mid-dim & Relativistic-$8M$ & $c=2,\,m=4.772$ & Cholesky & 0.737 & 306.5 & 1.29 \\
        Mid-dim & Relativistic-$4M$ & $c=2,\,m=2.386$ & Cholesky & 0.765 & 303.7 & 1.34 \\
        Mid-dim & Relativistic-$2M$ & $c=2,\,m=1.193$ & Cholesky & 0.798 & 381.3 & 1.00 \\
        Mid-dim & Relativistic & $c=2,\,m=0.597$ & Cholesky & 0.897 & 421.1 & 0.82 \\
        Mid-dim & Relativistic-$0.5M$ & $c=2,\,m=0.298$ & Cholesky & 0.976 & 770.6 & 0.32 \\
        Mid-dim & Relativistic-$4c$ & $c=8,\,m=0.597$ & Cholesky & 0.711 & 483.0 & 0.63 \\
        Mid-dim & Relativistic-$2c$ & $c=4,\,m=0.597$ & Cholesky & 0.774 & 446.4 & 0.57 \\
        Mid-dim & Relativistic-$0.5c$ & $c=1,\,m=0.597$ & Cholesky & 1.035 & 800.4 & 0.37 \\
        Mid-dim & \ST{} & $\nu=4$ & Cholesky & 1.033 & 526.0 & 0.66 \\
        Mid-dim & \ST{}-$0.75\nu$ & $\nu=3$ & Cholesky & 1.135 & 591.3 & 0.58 \\
        Mid-dim & \ST{}-$0.5\nu$ & $\nu=2$ & Cholesky & 1.261 & 713.4 & 0.48 \\
        Mid-dim & Gaussian & $m=1$ & Cholesky & 0.709 & 487.4 & 0.75 \\
        \cmidrule(l){4-7}
        Mid-dim & Relativistic & $c=2,\,m=0.597$ & Classic & 0.154 & 342.0 & 0.12 \\
        Mid-dim & \ST{} & $\nu=4$ & Classic & 0.311 & 1259.9 & 0.07 \\
        Mid-dim & Gaussian & $m=1$ & Classic & 0.135 & 242.4 & 0.15 \\
        \midrule
        High-dim & Relativistic-$2M$ &  $c=2,\,m=1.193$ & Cholesky & 0.942 & 822.9 & 0.00034 \\
        High-dim & Relativistic & $c=2,\,m=0.597$ & Cholesky & 1.013 & 1111.2 & 0.00025 \\
        High-dim & \ST{} & $\nu=4$ & Cholesky & 1.166 & 841.9 & 0.00034 \\
        High-dim & Gaussian & $m=1$ & Cholesky & 0.846 & 1119.7 & 0.00026 \\

        \midrule
    Lower-dim
    & Gaussian           & $m=1$ & Cholesky & 0.708 & 251.84 & 1.06 \\
    Lower-dim
    & Relativistic       & $c=2,\,m=0.597$ & Cholesky & 0.828 & 569.95 & 0.92 \\
    Lower-dim
    & \ST{}        & $\nu=4$ & Cholesky & 1.027 & 305.55 & 1.06 \\
    
    \cmidrule(l){4-7}

    Lower-dim
    & Gaussian           & $m=1$ & Classic  & 0.583 & 27.98  & 6.95 \\
    Lower-dim
    & Relativistic       & $c=2,\,m=0.597$ & Classic  & 0.725 & 19.95  & 7.97 \\
    Lower-dim
    & \ST{}        & $\nu=4$ & Classic  & 0.843 & 26.24  & 5.82 \\
        
    \bottomrule
    \end{tabular}
    \caption{Summary of results for the mid-dim, high-dim, and lower-dim experiments, which involve $12\,993$ parameters, $4\,202\,436$ parameters, and $225$ parameters, respectively. The rightmost three columns report the performance metrics: the Fraction of Missing Information (FMI; higher is better), the average correlation length $\hat{\tau}$ (lower is better), and the Effective Sample Rate ($\ESR{}$) averaged over the spectral variables (higher is better). The table also lists the experimental set, momentum distribution, parameterization, and associated parameters.}
    
    \label{tab:results}
\end{table*}

\subsection{Mid-dim set}

For the mid-dimensional `mid-dim' set of experiments, we produce full-sky galaxy clustering (spin-weight $0$ or $D$ field) and weak-lensing (spin-weight $2$ or $E,B$ modes) data maps with $12\,288$ \textsc{Healpix} \citep{Healpix} pixels ($N_\textrm{SIDE}=32$) in one tomographic bin. These maps are generated from a Gaussian realization of a cosmological power spectrum representative of expectations for Stage IV surveys \citep{desi_fullshape_2024, LSST, Euclid, Loureiro_2023}, and we restrict our analysis to multipoles in the range $\ell=4$ to $\ell=64$. To emulate realistic survey conditions, we add Poisson noise to the spin-weight~$0$ field (with a total object count of $445\,562\,734$) and white noise to the spin-weight~$2$ field (standard deviation $0.001$), following the setup used in \cite{Loureiro_2023}.

Running \almanac{} on these noisy inputs produces a posterior distribution, featuring a characteristic funnel structure in the high signal-to-noise regime, with $12\,627$ variables associated with the spherical harmonic coefficients and an additional $366$ variables for the power spectra ($DD$, $DE$, $DB$, $EE$, $EB$, and $BB$). We test the kinetic energy definitions introduced in Section~\ref{sec:kinetic_energies_with_bounded_velocity}, exploring a range of parameters (rest mass $m$ and speed of light $c$ for relativistic momenta, and degrees of freedom $\nu$ for \ST{}) as listed in Table~\ref{tab:results}. These choices are combined with both available parameterizations (`Classic' and `Cholesky'), allowing us to evaluate the interplay between momentum distribution, hyperparameters, and coordinate system.

\subsection{High-dim set}

For the high-dimensional `high-dim' set of experiments, we produce masked sky weak lensing (spin-weight $2$ or $E,B$ modes) data maps in two redshift bins, with $N_\textrm{SIDE}=512$ equivalent to $3\,145\,728$ Healpix pixels. As in the mid-dim case, these maps come from a Gaussian realization of a known cosmological power spectrum expected for Stage IV surveys, now from $\ell=4$ to $\ell=1024$, plus white noise (standard deviation of $0.001$). This setup, identical to the one in \cite{Loureiro_2023}, resembles a realistic cosmological application; it contains both high and low signal-to-noise regimes. 

Each `high-dim' experiment has $4\,202\,436$ variables associated with the spherical harmonic coefficients and $10\,210$ associated with the power spectra ($E_1E_1$, $E_1B_1$, $E_1E_2$, $E_1E_2$, $B_1B_1$, $B_1E_2$, $B_1B_2$, $E_2E_2$, $E_2B_2$, $B_2B_2$). For this set of experiments we only use the `Cholesky' parameterization, which had been revealed in the `mid-dimensional' set to have better performance than the `Classic' one (consistent with results in \cite{Loureiro_2023}). Due to the extremely high-dimensional problem of this set, we select only two different values of the rest mass parameter $m$, as detailed in Table~\ref{tab:results}, and only one value of the speed of light parameter $c$. For the \ST{} kinetic energy, we only select one value of the $\nu$ parameter.

\subsection{Lower-dim Set}

Additionally we include a lower-dimensional `lower-dim' set, motivated by the `mid-dim' configuration containing what is still a substantial number of variables ($12\,993$) and seeking to know therefore how the sampler behaves in a yet lower-dimensional regime.

To achieve this, we use the same input setup as in the `mid-dim' case but restrict the sampler to $\ell_{\mathrm{max}} = 8$, reducing the total number of variables to $225$ (including both the spherical harmonic coefficients and the power spectra). We implement the kinetic energy definitions described in Section~\ref{sec:kinetic_energies_with_bounded_velocity}, but due to the exploratory nature of this test, we focus only on the Gaussian, Relativistic, and \ST{} cases using their default parameter values. Each is tested under both the `Classic' and `Cholesky' parameterizations.

\section{Results}
\label{sec:results}

For each spectra variable $i$ in the different sets of experiments, we computed the correlation length $\hat{\tau}_i$ \citep{corr_length_estimator} and the Effective Sample Rate $\ESR{}_i$ as defined in Eq.~\ref{def_ESR}. The first is a measure of sampler performance (and thus of the momentum distribution being tested) while the second is a direct measure of sampling efficiency. We use $\ESR{}$ and $\hat{\tau}$ to denote these quantities averaged over all spectral variables, even when their distributions are not Gaussian. We also compute the median of those quantities across spectral variables, and find that it follows the same trend as the mean across experiments, confirming that the average remains a reliable metric. 

Table~\ref{tab:results} reports, for each experiment, the averaged Effective Sample Rate $\ESR{}$, the averaged correlation length $\hat{\tau}$, and the Fraction of Missing Information (FMI); it also lists the experimental set, coordinates, and parameters.

Table~\ref{tab:results} shows that the Cholesky parameterization consistently outperforms the Classic parameterization across all metrics. With other parameters held fixed, Cholesky yields higher FMI and higher $\ESR{}$ values, with an improvement in the latter by a factor of 5 to 10. For example, with Gaussian momentum in the mid-dim set, Cholesky yields FMI $= 0.709$ versus $0.135$ with Classic, and $\ESR{} = \qty{0.75}{\per\second}$ versus $\qty{0.15}{\per\second}$. As discussed in \cite{Loureiro_2023}, FMI values below $0.7$ indicate poor convergence; notably, all Classic runs fall into this category. Although some Classic runs exhibit slightly shorter correlation lengths, this apparent improvement is misleading.

Because wall-clock time is the only uncontrolled variable in Eq.~\ref{def_ESR}, and runs under the same parameterization should have comparable computational cost, we quantified timing variability. Across the Cholesky runs in the mid-dim set, the standard deviation of $t_{\mathrm{total}}$ was $119.40$ seconds. To assess hardware-induced fluctuations, we repeated one configuration ten times; the standard deviation of $t_{\mathrm{total}}$ for this configuration was $137.50$ seconds. These values are comparable, indicating that system-level variability is of the same order as within-parameterization timing differences. Importantly, the performance improvements observed, such as the 5–10$\times$ ESR gains for Cholesky over Classic, are far larger than this timing noise, confirming the robustness of our conclusions.

Among the relativistic momentum distributions, Relativistic-4M under Cholesky achieves the best average efficiency ($\ESR{} = \qty{1.34}{\per\second}$) and the best performance overall ($\hat{\tau} = 303.7$). This highlights the importance of carefully selecting parameters for each distribution. Recall that these distributions are expressed in multiples of the rest mass parameter, normalized under the unit variance criterion defined in Section~\ref{sec:sampling_from_momentum_distributions}.

Looking at FMI in the mid-dim experiments, we see substantially higher values for \ST{} and for relativistic distributions with lighter masses or lower light speeds. \ST{} consistently yields FMI $> 1$, peaking at $\nu = 2$ (FMI $= 1.261$). For relativistic momenta, decreasing the mass parameter $m$ increases FMI (from 0.737 at $m=4.772$ to 0.976 at $m=0.298$). Similarly, decreasing the light speed also increases FMI (from 0.711 at $c=8$ to 1.035 at $c=1$).

The Effective Sample Rate remains low across all mid-dim experiments, but we note a trade-off: higher FMI often coincides with lower $\ESR{}$. For example, Relativistic-$0.5M$ achieves high FMI (0.976) but low $\ESR{}$ ($\qty{0.32}{\per\second}$), whereas Relativistic-$2M$ achieves moderate FMI (0.798) but higher $\ESR{}$ ($\qty{1.00}{\per\second}$). This suggests that lighter-tailed momentum distributions improve mixing (FMI) but reduce raw sampling efficiency.

The high-dim cases are much higher dimensional than the mid-dim cases and hence are much more computationally expensive. Thus for the high-dim cases we restricted ourselves to the Cholesky parameterization, as this performed better in the mid-dim experiments. Given resource constraints, we also limited the set of kinetic energy definitions that we tested.

From Table~\ref{tab:results}, we see that, as expected, $\ESR{}$ in the high-dim sets drops dramatically (to $\sim \qty{e-4}{\per\second}$), compared with the mid-dim experiments. This reduction arises because the dimensionality increases by 2.5 orders of magnitude, so the sampler must generate the same number of samples in a much more computationally expensive setting. This explains the differences in $\ESR{}$, but not in correlation length, compared to mid-dim. Importantly, FMI and correlation lengths remain informative. \ST{} ($\nu=4$) again yields the highest FMI (1.166), consistent with the mid-dim case. The Relativistic distributions (especially Relativistic-$2M$) achieve competitive correlation lengths. As before, Gaussian gives the lowest FMI (0.846).

\begin{figure}
    \centering
    \includegraphics[width=0.9\linewidth]{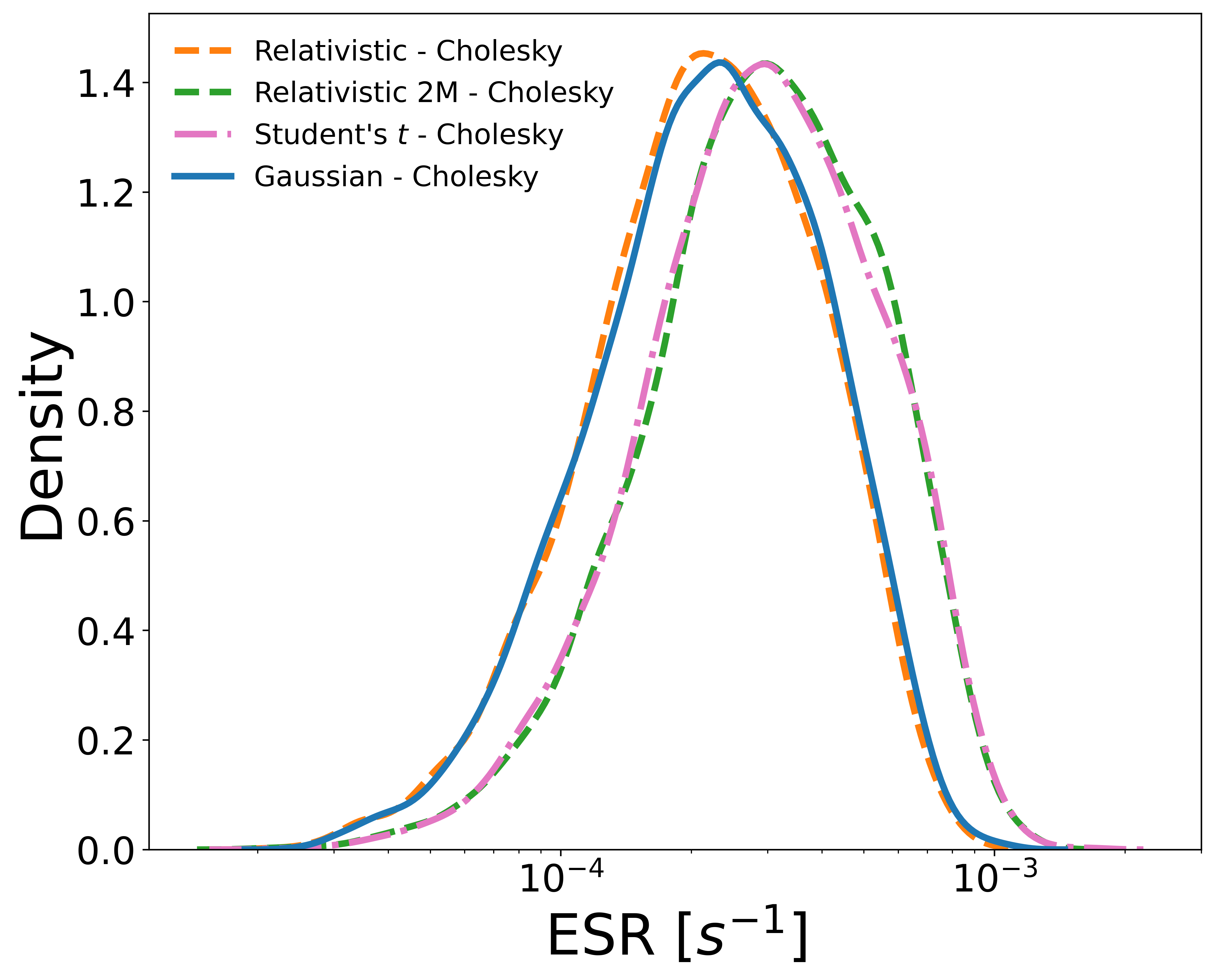}
    \caption{
    The Effective Sample Rate depends on the parameter considered.  Here we plot the distributions of the spectrum variables for the `high-dimensional' set of experiments in Section~\ref{sec:almanac_experiments}. The colours and markers indicate the different kinetic energy definitions. The $x\text{-axis}$ is log-spaced and the $y\text{-axis}$ shows the probability density of this logarithmic variable. See Section~\ref{sec:results} for a discussion of the `two pairs' appearance of this plot.}
    \label{fig:ESR_kde}
\end{figure}

The density plot in Fig.~\ref{fig:ESR_kde} supports these observations, where the $\ESR{}$ distributions for Relativistic, Relativistic-2M, \ST{}, and Gaussian are closely aligned, peaking around $10^{-4}$ s$^{-1}$, but with subtle differences in width and skewness. Relativistic-2M and \ST{} have slightly broader $\ESR{}$ distributions, consistent with a more robust exploration of the posterior.

\begin{figure*}
    \centering
    \includegraphics[width=\textwidth]{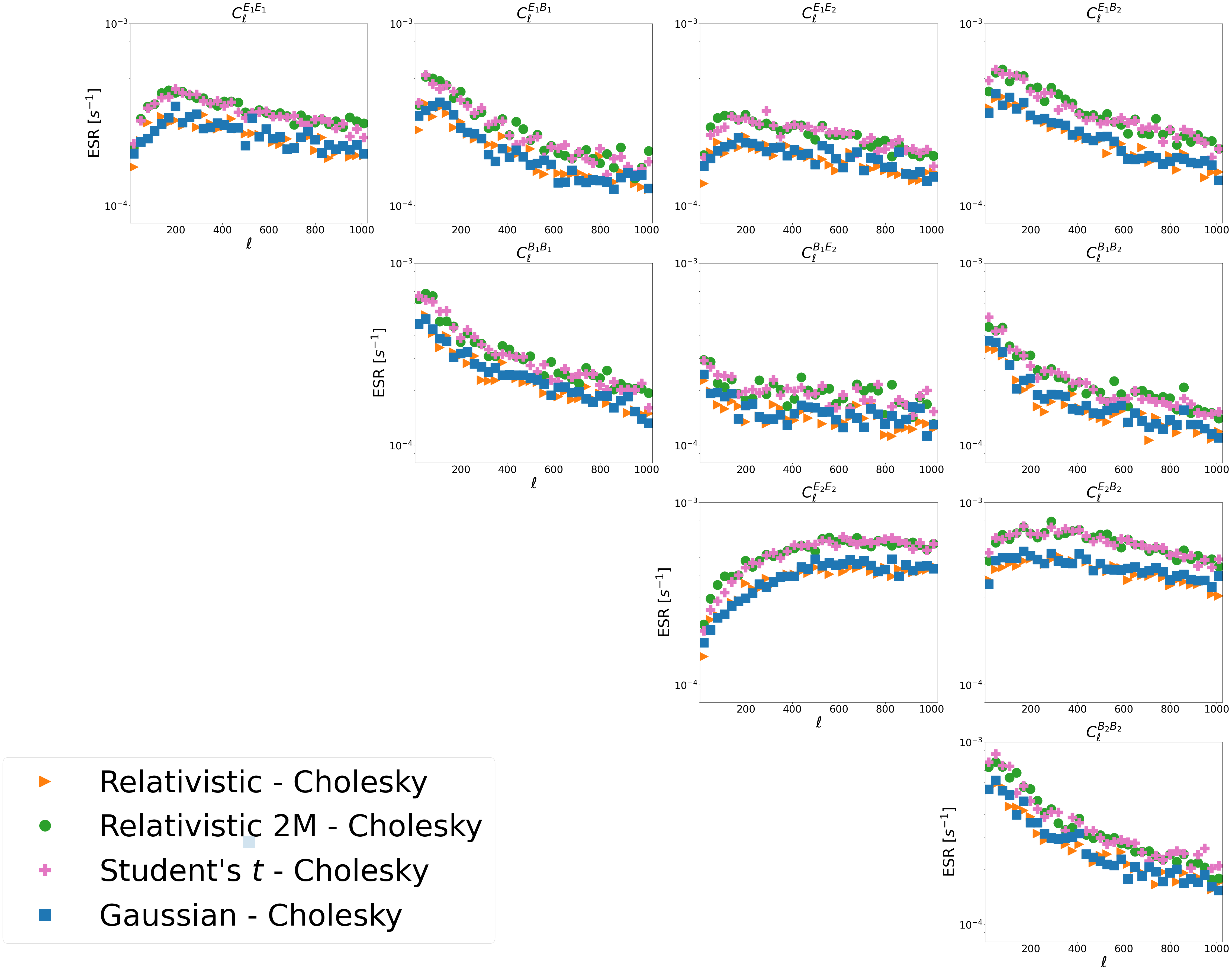}
    \caption{Effective Sample Rate as function of multipole moment $\ell$, ordered according to the different power spectra in the two redshift bins for the high-dim case. The colours and markers indicate the different kinetic energy definitions used in the sampler. Note how the efficiency of the different spectra changes according to $\ell$, especially those containing $B-$modes. For clarity, only a subset of $\ell$ values between 4 and 1024 is displayed, plotted every 30 modes.}
    \label{fig:corr_length_cls_spin2_2bins_signal}
\end{figure*}

Fig.~\ref{fig:corr_length_cls_spin2_2bins_signal} shows how, in the high-dim case, the per-$\ell$ values for $\ESR{}$ vary with $\ell$ for the various spectral components ($E_1E_1$, $E_1B_1$, etc.), for the various kinetic energy definitions. It appears that each spectral component has a basic shape for the $\ESR{}$ as a function of $\ell$; for example, the plots for $E_2E_2$ appear to be convex down, with a maximum around $\ell = 600$. Changing the kinetic energy definition appears then simply to scale each basic shape by a factor that is independent of spectral component. We hypothesize that these basic shapes are determined purely by the shape of the posterior. As a result, a kinetic energy definition that performs well appears to do so by doing relatively better on each parameter equally (rather than by being better on some parameters but not on others). As shown in Table~\ref{tab:results}, Relativistic and Gaussian happen to have a similar overall efficiency, as do Relativistic-2M and \ST{}; this results in the plots in Fig.~\ref{fig:corr_length_cls_spin2_2bins_signal} having their `two pairs of two lines' appearance.
Fig.~\ref{fig:ESR_kde} then summarises these values into a single histogram, which has a similar appearance (but smoother as noise has been averaged away). Further work will be required to determine if this explanation is correct.

To conclude: in the high-dim case, relativistic momentum distributions do not yield significantly better performance than Gaussian. This motivates further exploration of alternative rest mass parameter values. Following the mid-dim results, we increased the rest mass parameter, running tests at double its previous value. These findings emphasize the importance of self-tuning hyperparameters and the need for more careful selection of kinetic energy hyperparameters, as in \cite{Selftuning}.

For completeness, we also consider a lower-dimensional variant of the mid-dim setup, containing only 225 variables and operating in a very high signal-to-noise regime, described in Section~\ref{sec:almanac_experiments}. In this setting the sampler behaves differently: the Classic parameterization now clearly outperforms the Cholesky parameterization. Correlation lengths under Classic are an order of magnitude smaller (e.g.\ $\hat{\tau}\!\sim\!28$ versus $\hat{\tau}\!\sim\!252$ for Gaussian momentum, and $\hat{\tau}\!\sim\!20$ versus $\hat{\tau}\!\sim\!570$ for relativistic momentum), and the $\ESR{}$ are correspondingly higher, reaching $5.8-\qty{8.0}{\per\second}$ compared to $\sim \qty{1}{\per\second}$ for Cholesky. The FMI values follow the same pattern: although Gaussian–Classic gives FMI $=0.583<0.7$, both relativistic–Classic and \ST{}–Classic achieve FMI above the convergence threshold. This reversal in favour of the Classic coordinates in the lower-dim set, high–signal-to-noise regime is consistent with established results on centered versus non-centered parameterizations \citep{gorinova2019automaticreparameterisationprobabilisticprograms}, where centered forms perform better when parameters are strongly data-informed \citep[e.g.][]{betancourt2013hamiltonianmontecarlohierarchical, kingma2022autoencodingvariationalbayes, JSSv076i01}. However, this regime is not representative of our main cosmological applications, which high-dimensional and contain both high and low signal-to-noise regions. The overall trends and conclusions drawn from the mid-dim and high-dim experiments therefore remain unchanged.

\section{Discussion}
\label{sec:discussion}

The results presented in Section~\ref{sec:results} show that gains in sample rate are possible not only by tuning the parameters of the momentum (kinetic energy) distribution during the early stages of sampler design, in agreement with the results in \cite{Loureiro_2023,Sellentin_2023,Selftuning,livingstone2019kinetic}, but also by the choice of the form of the momentum distribution. These choices and early optimization affect convergence, efficiency, and computational cost, particularly in models with high dimensionality. From our experiments, we conclude the following:

\begin{itemize}
    \item The Cholesky parametrization outperformed the Classic parametrization across all metrics. This confirms again that the parametrization has a fundamental impact on the ability of the sampler to explore the posterior. In fact, runs under the Classic parameterization not only produced lower FMI values, meaning a poor convergence, but also misleadingly shorter correlation lengths (except in the lower-dim, high–signal-to-noise regime of the lower-dim set). This reinforces the point that apparent efficiency can mask poor convergence.

    \item The choice of momentum distribution introduces a clear trade-off between efficiency and robustness. Heavy-tailed distributions, such as \ST{} or relativistic distributions with small mass parameters, improve mixing (higher FMI, lower correlation length), but at the expense of raw sampling speed (lower ESR). Contrariwise, distributions with moderate tails, such as relativistic-$2M$, offer better ESR at the cost of slightly reduced mixing. 
    
    \item We found a scalability challenge for the high-dimensional case. The Effective Sample Rate dropped by several orders of magnitude compared to the mid-dimensional tests. This effect originates directly from the increase in dimensionality, which amplifies the computational cost per iteration. However, the relative performance of different parameterizations and momentum distributions remained consistent with the mid-dimensional case, suggesting that insights gained from mid-dimensional experiments can meaningfully guide choices in more complex settings.
    
    \item These findings emphasize the need for hyperparameter tuning. For relativistic distributions, the rest mass and speed of light parameters strongly affect both FMI and ESR. Our exploratory tests with modified rest mass indicate that careful tuning can significantly improve performance. This aligns with recent work on self-tuning strategies \citep{Selftuning}, which may provide a methodical way to optimize sampler settings without extensive trial-and-error.
\end{itemize}

In summary, our study shows that sampler design must be tailored to the problem at hand rather than relying on a universal choice. While performance gains from modifying the parameterization and kinetic energy are possible, they are generally modest and strongly problem-dependent. For high-dimensional problems, these decisions become even more critical, as computational costs grow rapidly. Future work should therefore focus on developing adaptive methods that automatically tune momentum distributions and parameterizations to the form of the posterior distribution. Such methods could provide a route toward self-optimizing samplers that maintain performance across diverse models and dimensional regimes.

\section*{Acknowledgments}

We thank the support staff of Leiden University's ALICE High Performance Computing infrastructure. We thank Daniela Grand\'on, Zeynab Ashuri, and Mariia Marinichenko for useful discussions. The work presented here was made possible thanks to the following software: \almanac{} \citep{Loureiro_2023,Sellentin_2023}, \textsc{HEALPix} \citep{Healpix}, \textsc{libsharp2} \citep{2013-Libsharp,2018-Libsharp2}, \textsc{Seaborn} \cite{Waskom2021}, \textsc{matplotlib} \citep{Hunter:2007-matplotlib}, and \textsc{NumPy} \citep{harris2020numpy}.

\bibliographystyle{mnras}
\bibliography{Bib}
\end{document}